\pdfoutput=1
\documentclass[letterpaper,pdftex]{article}
\usepackage[margin=1.5in]{geometry}
\usepackage[sort&compress]{natbib}

\let\oldtitle\title
\renewcommand\title[1]{%
    \hypersetup{pdftitle={#1}}%
    \oldtitle{#1}%
    \def\thetitle{#1}%
    \pdfbookmark[0]{#1}{title}}

\newcommand\andnext{\unskip, }%
\newcommand\authoraffils[1]{\mbox{}\\\mbox{}\\#1}
\newcommand\corremail[1]{\mbox{}\\\emaillink{#1}}
\let\oldauthor\author
\renewcommand\author[1]{%
    \begingroup
        \providecommand\and{}%
        \renewcommand\and{and }%
        \providecommand\andnext{}%
        \renewcommand\andnext{and }%
        \providecommand\textsuperscript[1]{}%
        \renewcommand\textsuperscript[1]{}%
        \providecommand\authoraffils[1]{}%
        \renewcommand\authoraffils[1]{}%
        \providecommand\corremail[1]{}%
        \renewcommand\corremail[1]{}%
        \providecommand\thanks[1]{}%
        \renewcommand\thanks[1]{}%
        \hypersetup{pdfauthor={#1}}%
        \xdef\theauthor{#1}%
    \endgroup
    \oldauthor{#1}%
    }

\usepackage{natbib}
\usepackage[shortcuts]{extdash}
\usepackage{todonotes}
\usepackage{verbatim}

\usepackage{subcaption}
\usepackage{rotating}
\usepackage{floatpag}
\rotfloatpagestyle{empty}

\usepackage{mathtools}    
\usepackage{amssymb}
\usepackage{amsthm}
\usepackage{varioref}

\labelformat{subfigure}{\thefigure\textup{(#1)}}
\labelformat{equation}{\textup{(#1)}}

\usepackage[%
        pagebackref=false,
        bookmarks=true,bookmarksopen=true,
        bookmarksnumbered=true,
        breaklinks=true,
        colorlinks=true,
        anchorcolor=blue,citecolor=blue,
        urlcolor=blue,linkcolor=blue,filecolor=blue,
        menucolor=blue,
    ]{hyperref}

\let\orgautoref\autoref
\providecommand{\Autoref}
        {\def\equationautorefname{Equation}%
         \def\figureautorefname{Figure}%
         \def\sectionautorefname{Section}%
         \def\appendixautorefname{Appendix}%
         \def\Itemautorefname{Item}%
         \def\tableautorefname{Table}%
         \def\thmautorefname{Theorem}%
         \def\lemautorefname{Lemma}%
         \def\propautorefname{Proposition}%
         \def\asmpautorefname{Assumption}%
         \def\corautorefname{Corollary}%
         \def\defnautorefname{Definition}%
         \def\conjautorefname{Conjecture}%
         \def\exmpautorefname{Example}%
         \def\remautorefname{Remark}%
         \def\footnoteautorefname{Note}%
         \def\caseautorefname{Case}%
         \orgautoref}

\providecommand{\Autorefs}
        {\def\equationautorefname{Equations}%
         \def\figureautorefname{Figures}%
         \def\sectionautorefname{Sections}%
         \def\appendixautorefname{Appendices}%
         \def\Itemautorefname{Items}%
         \def\tableautorefname{Tables}%
         \def\thmautorefname{Theorems}%
         \def\lemautorefname{Lemmas}%
         \def\propautorefname{Propositions}%
         \def\asmpautorefname{Assumptions}%
         \def\corautorefname{Corollaries}%
         \def\defnautorefname{Definitions}%
         \def\conjautorefname{Conjectures}%
         \def\exmpautorefname{Examples}%
         \def\remautorefname{Remarks}%
         \def\footnoteautorefname{Notes}%
         \def\caseautorefname{Cases}%
         \orgautoref}

\renewcommand{\autoref}
        {\def\equationautorefname{Equation}%
         \def\figureautorefname{Figure}%
         \def\sectionautorefname{Section}%
         \def\appendixautorefname{Appendix}%
         \def\Itemautorefname{item}%
         \def\tableautorefname{Table}%
         \def\thmautorefname{Theorem}%
         \def\lemautorefname{Lemma}%
         \def\propautorefname{Proposition}%
         \def\asmpautorefname{Assumption}%
         \def\corautorefname{Corollary}%
         \def\defnautorefname{Definition}%
         \def\conjautorefname{Conjecture}%
         \def\exmpautorefname{Example}%
         \def\remautorefname{Remark}%
         \def\footnoteautorefname{Note}%
         \def\caseautorefname{Case}%
         \orgautoref}

\providecommand{\autorefs}
        {\def\equationautorefname{Equations}%
         \def\figureautorefname{Figures}%
         \def\sectionautorefname{Sections}%
         \def\appendixautorefname{Appendices}%
         \def\Itemautorefname{items}%
         \def\tableautorefname{Tables}%
         \def\thmautorefname{Theorems}%
         \def\lemautorefname{Lemmas}%
         \def\propautorefname{Propositions}%
         \def\asmpautorefname{Assumptions}%
         \def\corautorefname{Corollaries}%
         \def\defnautorefname{Definitions}%
         \def\conjautorefname{Conjectures}%
         \def\exmpautorefname{Examples}%
         \def\remautorefname{Remarks}%
         \def\footnoteautorefname{Notes}%
         \def\caseautorefname{Cases}%
         \orgautoref}

\usepackage{graphicx}
\usepackage[export]{adjustbox}  

\usepackage{makeidx}
\makeindex

\title{Noise and Function}
\author{Steven Weinstein%
            \thanks{Department of Philosophy,
                University of Waterloo, Waterloo, ON, Canada N2L 3G1}%
            \and%
            Theodore P.~Pavlic%
            \thanks{School of Computing, Informatics, and Decision Systems
                Engineering, Arizona State University, Tempe, AZ, USA 85287}%
        }
\date{}

\begin{document}
\maketitle

\begin{abstract}
Noise is widely understood to be something that interferes with a signal
or process. Thus, it is generally thought to be destructive, obscuring
signals and interfering with function. However, early in the 20th
century, mechanical engineers found that mechanisms inducing additional
vibration in mechanical systems could prevent sticking and hysteresis.
This so\-/called ``dither'' noise was later introduced in an entirely
different context at the advent of digital information transmission and
recording in the early 1960s. Ironically, the addition of noise allows one
to preserve information that would otherwise be lost when the signal or
image is digitized. As we shall see, the benefits of added
noise in these contexts are closely related to the phenomenon which has
come to be known as \emph{stochastic resonance}, the original version of
which appealed to noise to explain how small periodic fluctuations in
the eccentricity of the earth's orbit might be amplified in such a way
as to bring about the observed periodic transitions in climate from ice age to
temperate age and back. These noise-induced transitions have since been invoked to explain a
wide array of biological phenomena, including the foraging and tracking
behavior of ants. Many biological phenomena, from foraging to gene
expression, are noisy, involving an element of randomness. In this
paper, we illustrate the general principles behind dithering and stochastic
resonance using examples from image processing, and then show how the
constructive use of noise can carry over to systems
found in nature.
\end{abstract}

\section{Introduction}
\label{sec:intro}

We are surrounded by noise. The controlled explosions of internal
combustion engines combine with the roar of rubber on asphalt to create
the drone of road and highway traffic. Weed\-/whackers, lawnmowers, and
leaf blowers can turn sunny suburban summer days into a buzzing
confusion. Indeed, the entire universe is filled with the faint din that
is the
\index{CMB|see{cosmic microwave background radiation}}\index{cosmic microwave background radiation}%
cosmic microwave background~(CMB) radiation, leftover electromagnetic
radiation from the \index{big bang}%
The CMB was discovered when \citet{PW65} went looking for the source of
annoying hiss plaguing their new radio telescope, hiss that threatened
to obscure signals from distant stars and galaxies.  Noise seems to be entirely
destructive, thus something to be eliminated if possible.

Noise can be beneficial, however, in at least two ways. One is familiar,
the other paradoxical and far less well known. The familiar way is
simply as a source of variety. For example, genes undergo random mutation from
processes both external to the organism (e.g. cosmic rays) and internal~\citep{dobrindt2001whole}. 
This genotypic variation is the
source of heritable variation in phenotype, which is of course essential for the process of natural selection
~\citep{wagner2014arrival}. Phenotype can even vary within
isogenic populations due to variation in gene
expression~\citep{fraser2009chance}. This \emph{phenotypic noise}\index{phenotypic noise} is thought to
provide an evolutionary advantage for some microorganisms, as it
increases the chance that some will survive under stressful conditions.
The CMB noise, though destructive from the standpoint of the users of
radio telescopes, plays a constructive role in generating the tiny
variations in energy density in the early universe that are the seeds of
structure formation. The random fluctuations we call noise give rise to stars and galaxies and
galactic clusters. This much is familiar, at least to those working in
the relevant areas of biology or astrophysics.

Considerably less familiar is the role that noise can play in nonlinear systems,
in particular systems with one or more thresholds, points at which small
differences in input give rise to disproportionate differences
in output. Converting an analog signal into a digital signal involves
sampling the signal at regular intervals and writing down a digital approximation of the amplitude at each point.
For example, if one has a one-bit digital system with two possible values to represent the interval from 0 to 1, then there will be a threshold \index{threshold}at the analog value 0.5, below which any value will be digitally recorded as 0, and above which any value will be recorded as 1. More bits simply mean more thresholds, more ways to cut up the interval into discrete chunks. Neurons behave like single threshold
devices, firing when and only when the voltage across the cell membrane
reaches a certain activation threshold. What noise can do in a
threshold system is push the signal over the threshold, but in a way
very much unlike an amplifier.  Amplifiers multiply the signal, whereas
noise is additive. The implications and applications of this nonstandard
amplification are both deep and wide.  Here we will lay out as simply as
possible the principles behind this sort of noise benefit and then
illustrate its application.

\section{Shades of gray: Noise in image processing}
\label{sec:dithering}

\index{photography}Photographs are never veridical.  The information
coming through the lens is inevitably greater than the information
stored on the recording medium. A \index{photography!digital}digital
camera sensor has a finite \index{spatial resolution}spatial resolution;
the camera's sensor consists of a matrix of smaller individual sensors
corresponding to a single \index{pixel}``pixel'' of the image. Any
features of the image smaller than an individual pixel will be lost.
Associated with each pixel is a color. The color spectrum in the real
world is continuous, but the digital encoding of color is discrete, so
that in general, the color stored will only be an approximation of the
actual color. In other words, color information must be
\index{rounding error}\index{quantization noise!rounding|see{rounding error}}%
rounded off in order to be stored as a number on a digital computer. The
number of binary places available for each number is referred to as the
bit depth.

\begin{figure}\centering
    \subcaptionbox{\label{fig:Grayscale256}8 bit grayscale palette (256 shades)}[0.45\textwidth]
       {\includegraphics[height=0.4\textwidth, frame]{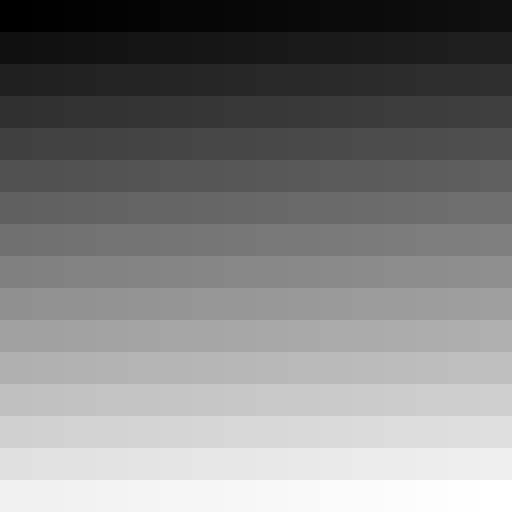}}
     \subcaptionbox{\label{fig:GrayscaleLena}8 bit Lena (256 shades)}[0.45\textwidth]
       {\includegraphics[height=0.4\textwidth, frame]{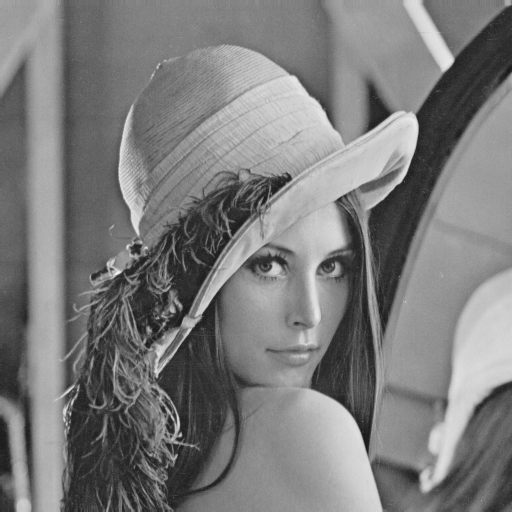}}

     \subcaptionbox{\label{fig:FixedThreshold}1 bit grayscale (2 shades)}[0.45\textwidth]
       {\includegraphics[height=0.4\textwidth, frame]{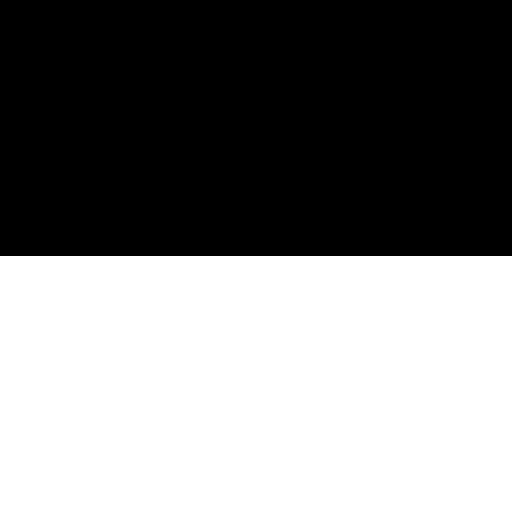}}
     \subcaptionbox{\label{fig:FixedThresholdLena}1 bit Lena (2 shades)}[0.45\textwidth]
       {\includegraphics[height=0.4\textwidth, frame]{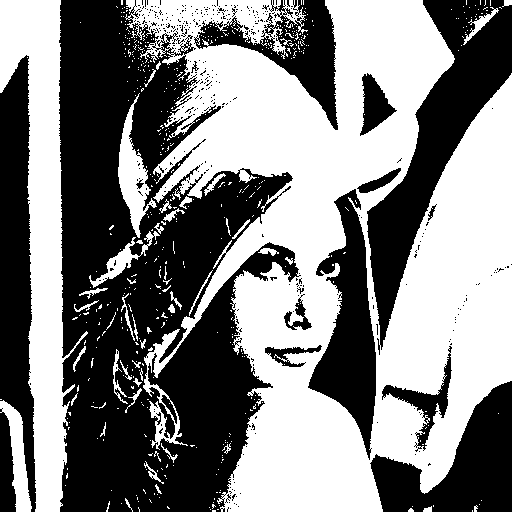}}

     \subcaptionbox{\label{fig:FixedThreshold8}3 bit grayscale (8 shades)}[0.45\textwidth]
       {\includegraphics[height=0.4\textwidth, frame]{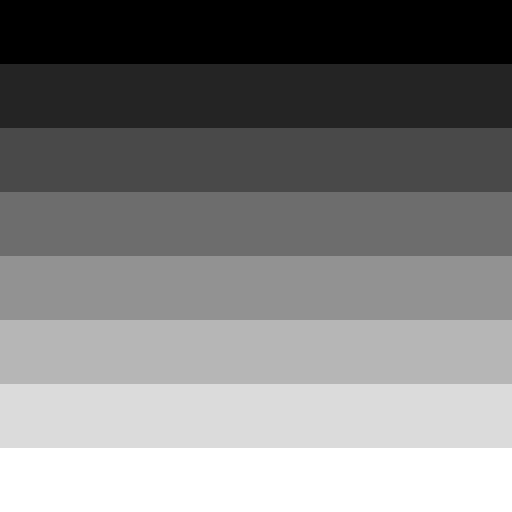}}
     \subcaptionbox{\label{fig:FixedThresholdLena8}3 bit Lena (8 shades)}[0.45\textwidth]
       {\includegraphics[height=0.4\textwidth, frame]{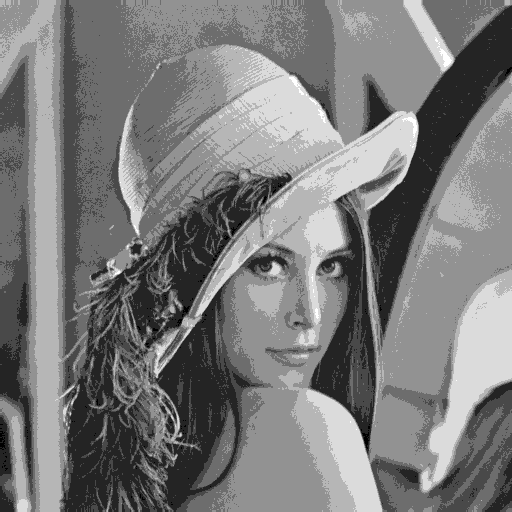}}
  \caption{Shades of gray}
\end{figure}

Suppose we have a digital image consisting of~7 megapixels.  Let's
consider a ``black and white" camera for simplicity, so that the colors
are shades of gray.  Each pixel has an~8 bit number attached to it
indicating the shade of gray, with 0 ($00000000$) corresponding (by
convention) to black, and $255$ ($11111111$) corresponding to white.
There are a total of $2^8 = 256$ shades of gray.\index{grayscale}
\Autoref{fig:Grayscale256} shows the palette, alongside an image of a
woman known as \index{Lena}Lena~\citep{Hutchison01}, rendered using this
palette (\autoref{fig:GrayscaleLena}).

Now, suppose we want to print these images. Indeed, you may well be
reading a printed version of this page, printed on a laser printer
capable of printing~300 dots per inch~(dpi). That resolution gives us
around~7 million evenly spaced points on a typical sheet of paper, so if
we were to use up an entire page to print the 7~megapixel image, we
would have a one-to-one correspondence between pixels and dots. If the
printer could print~256 shades of gray at any given point, then we'd
have perfect reproduction of the stored image. But the printer is not
nearly that flexible. At each dot location, most printers can print
either a black dot or nothing. Because the number of pixels and the
number of dots are approximately the same (in our example), we are
effectively reducing an~8-bit (per pixel) image to a~1-bit (per dot)
image; 256 shades of gray at each point get mapped to either
\index{black and white}black or white.

The obvious way to map the shades of gray is to impose a threshold  \index{threshold}as before,
whereby we print a black dot at a point if the corresponding pixel is
more than~50\% gray (numbers between~0 and~127), and we
otherwise leave it blank~(white) (numbers between~128 and~255). For an
image that has an equal distribution of lighter and darker grays, this
might appear to be as good as one can do. But a quick glance at
\autorefs{fig:FixedThreshold} and~\ref{fig:FixedThresholdLena} shows
the limitations of this simple thresholding method; an enormous amount of detail is lost. Increasing the
number of thresholds to create~8 shades of gray, as in
\autorefs{fig:FixedThreshold8} and~\ref{fig:FixedThresholdLena8}, yields
a noticeable improvement, but a printer that can only print in black and
white is limited to the performance of the~2 shade case.

\begin{figure}\centering
    \subcaptionbox{\label{fig:halftone_array}Halftone array}
        {\includegraphics[width=0.49\textwidth]{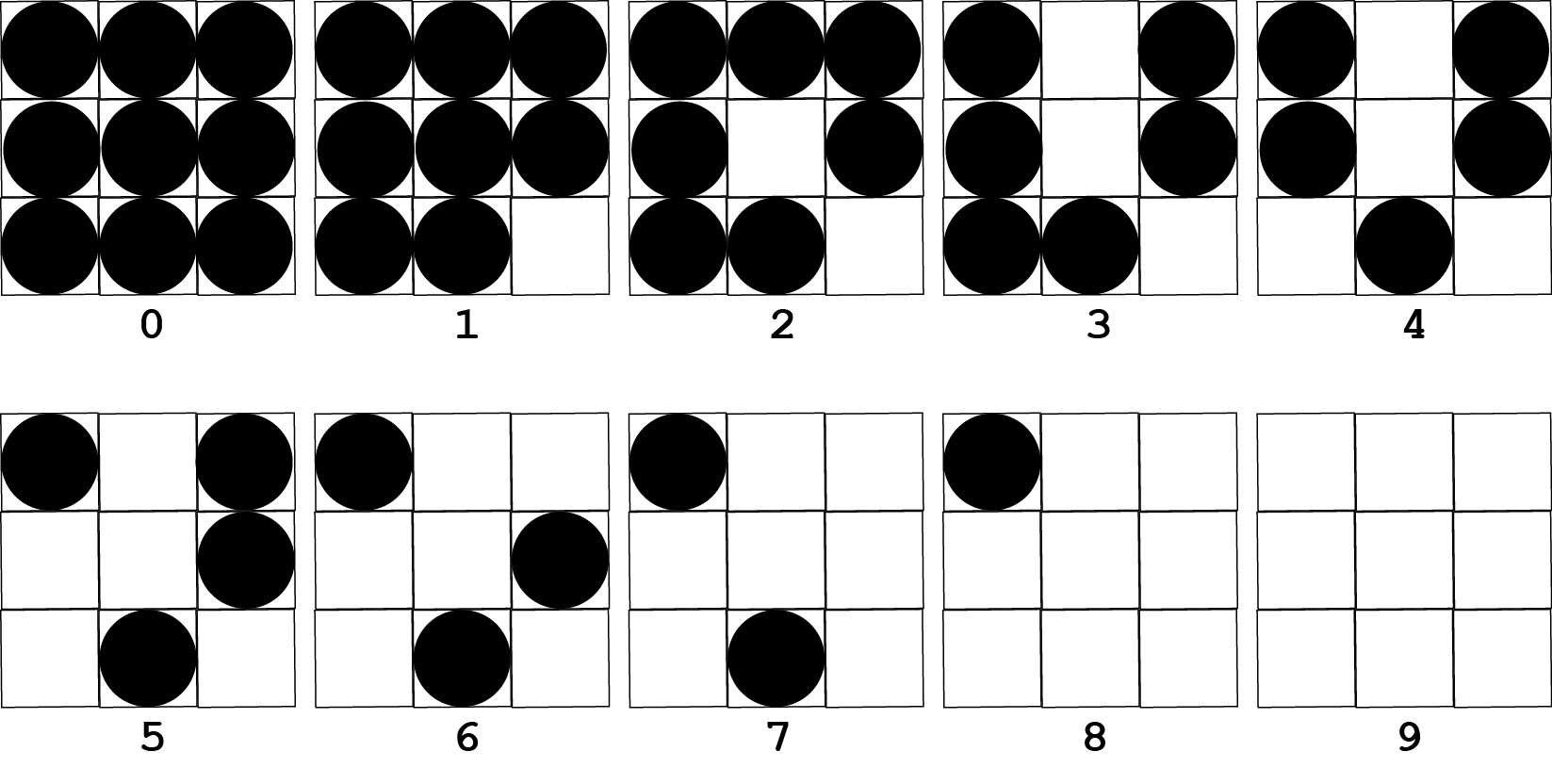}}
    \subcaptionbox{\label{fig:halftone_grayscale}Halftone grayscale palette}
        {\includegraphics[width=0.49\textwidth, frame]{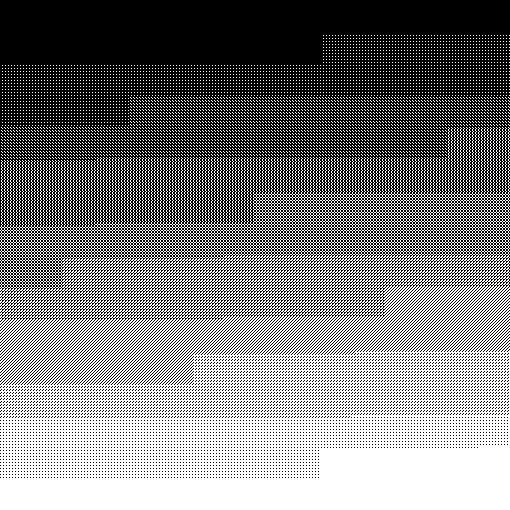}}
    \caption{Grayscale using halftone}
    \label{fig:halftone}
\end{figure}

However, there are clever methods to improve the fidelity of
black-and-white image reproductions. Traditional printed newspapers used
varying dot size to represent darker and lighter portions of an image.
Applying this \index{halftoning}\emph{halftone} concept to a device like
a laser printer or an LCD display with \emph{fixed} dot or pixel size
involves representing gray by varying the density of the distribution of
black dots in an array. Using $3 \times 3$ arrays of dots, we can
represent ten different shades of gray shown in
\autoref{fig:halftone_array}, which allows for a grayscale palette like
that shown in \autoref{fig:halftone_grayscale}.
A 300 dpi~(dots per inch) printer can print $100 \times 100$ patches of
gray per inch, where each patch has a $3 \times 3$ pixel area. Thus,
with a three-fold reduction in the effective spatial resolution of the
image, the number of shades which can be represented is increased from
two shades to ten. The same technique could be applied to $4 \times 4$
arrays of dots to achieve seventeen shades of gray at the cost of
further decreasing the spatial resolution.

\begin{figure}\centering
    \subcaptionbox{\label{fig:noisy_grayscale}8~bit grayscale (256 shades) w/~8 bit noise}
        {\includegraphics[width=0.45\textwidth, frame]{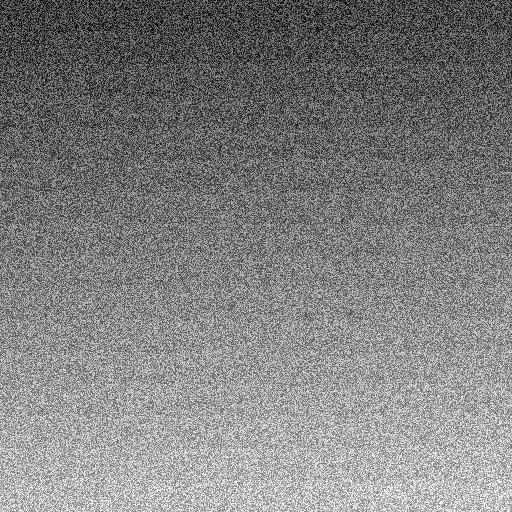}}
    \subcaptionbox{\label{fig:noisy_lena}8~bit Lena (256 shades) w/~8 bit noise}
        {\includegraphics[width=0.45\textwidth, frame]{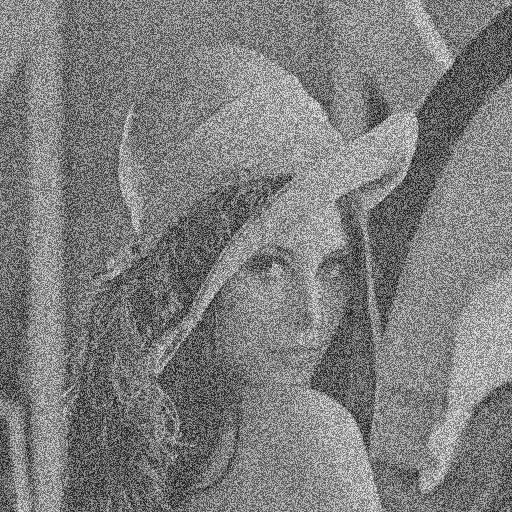}}

    \subcaptionbox{\label{fig:random_2level_grayscale}1~bit grayscale (2 shades)}
        {\includegraphics[width=0.45\textwidth, frame]{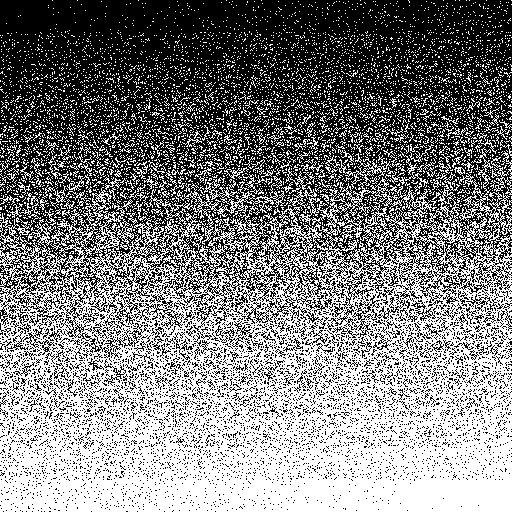}}
    \subcaptionbox{\label{fig:random_2level_Lena}1~bit Lena (2 shades)}
        {\includegraphics[width=0.45\textwidth, frame]{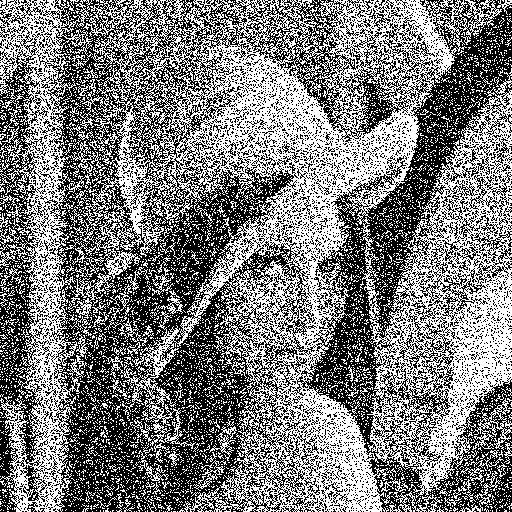}}
    \caption
        [Reduction from 256 shades (8~bits) to 2 shades (1~bit) using 8~bit random dither noise]
        {Reduction from 256 shades (8~bits) to 2 shades (1~bit) using 8~bit random dither noise.}
    \label{fig:2level8bitnoise}
\end{figure}

Although deterministic methods like \index{halftoning}halftoning can be
effective ways of trading spatial resolution for color resolution, they
introduce noticeable artifacts. Note the abruptness of the shifts in the
halftone grayscale palette in \autoref{fig:halftone_grayscale}. To avoid
this blockiness while still being able to trade spatial resolution for
color resolution, a very different approach can be used based on the
appropriate addition of random variation to pixel values. For each
pixel, we take the original grayscale value and add a random number
between 0 and 255.  An image made up of these random values would look
like visual ``noise''~-- it is a distribution of dots in arbitrary shades
of gray. The result of adding this noise to the image is an array of pixels with values between~0 and~510.
We can turn the resulting array back into an image by dividing
the values by two, thereby restoring the original range of 0~(black) to
255~(white). \Autorefs{fig:noisy_grayscale} and~\ref{fig:noisy_lena}
show the result: noisy versions of the original grayscale palette and
the original image.
This of course is not an improvement over the non-noisy 8-bit grayscale
image with 256 shades of gray. The noise does what we generally expect
noise to do: it degrades the image.

But recall that we added the noise not to improve the grayscale image
but to get a better result when we subsequently impose a threshold at~127 and convert to a black and white~(2~shade)
image. Imposing the threshold, we map any pixel with value~128 to white~(255), and any
pixel~127 or below to black~(0). The resulting pseudo\-/grayscale palette that
looks like \autoref{fig:random_2level_grayscale}, while the
resulting pseudo\-/grayscale Lena looks like \autoref{fig:random_2level_Lena}. The
results are instructive. The grayscale palette looks better than the
version in which noise was not imposed before thresholding
(\autoref{fig:FixedThreshold}), giving the impression of a variety of shades of gray.  Lena, however, does not look very good by
comparison with \autoref{fig:FixedThresholdLena}. The reason, as we
noted above, is that given enough pixels, the ratio of black dots to
white will closely approximate the degree of grayness of the original
shade from the~256-color palette. For a large number of pixels, observed
at a sufficiently great distance, we get an excellent representation of
gray. The problem with the Lena image is not that she has more
shades of gray, but that the shades tend to change over the scale of a
few pixels. Images of this sort are better treated by more sophisticated
techniques such as the Floyd\--Steinberg error diffusion
method~\citep{FS76}.

\begin{figure}\centering
   \subcaptionbox{\label{fig:noisy_grayscale32}8~bit grayscale (256 shades) w/~3 bit noise}
        {\includegraphics[width=0.45\textwidth, frame]{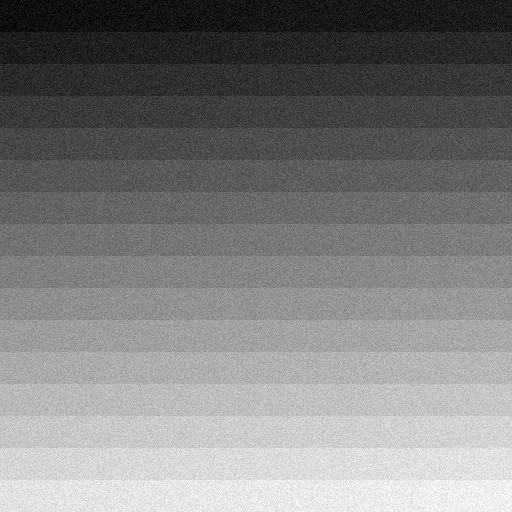}}
    \subcaptionbox{\label{fig:noisy_lena32}8~bit Lena (256 shades) w/~3 bit noise }
        {\includegraphics[width=0.45\textwidth, frame]{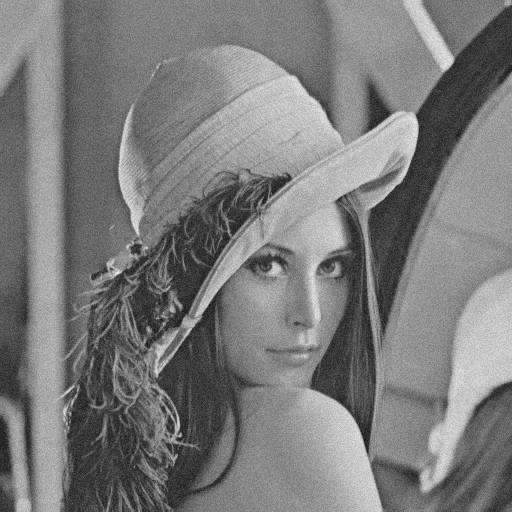}}

    \subcaptionbox{\label{fig:random_8level_grayscale}3 bit grayscale (8 shades)}
        {\includegraphics[width=0.45\textwidth, frame]{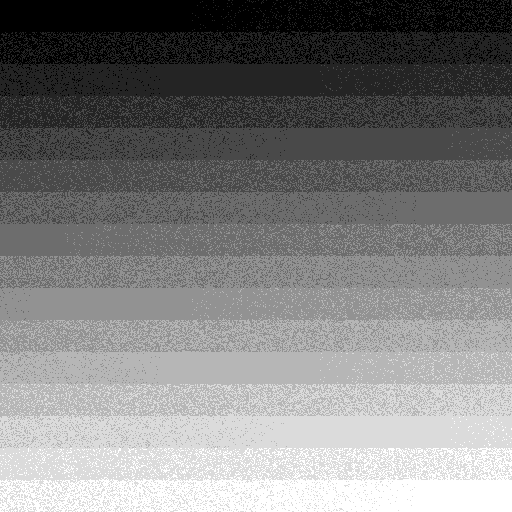}}
    \subcaptionbox{\label{fig:random_8level_Lena}3 bit Lena (8 shades)}
        {\includegraphics[width=0.45\textwidth, frame]{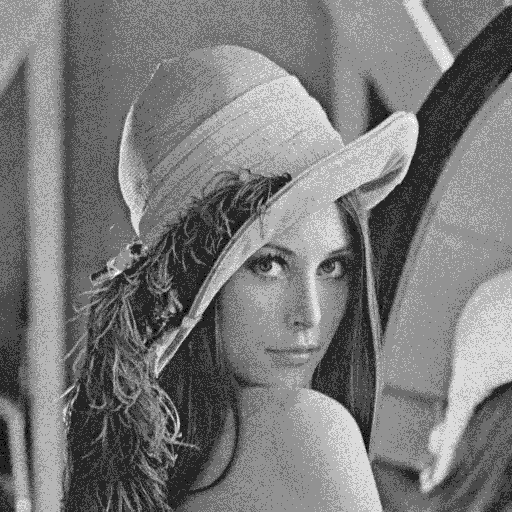}}    
    \caption
        [Reduction from 256~shades (8~bits) to 8~shades (3~bits) using 3~bit random dither noise (randomizing the 3 least significant bits (LSB)]
        {Reduction from 256~shades (8~bits) to 8~shades (3~bits) using 3~bit random dither noise (randomizing the 3 least significant bits (LSB).}
    \label{fig:2level1bitnoise}
\end{figure}

However, if we avail ourselves of~8 shades of gray (3 bits) rather than just
black and white (1 bit), the use of random noise is much more effective.
\autoref{fig:noisy_grayscale32} and \autoref{fig:noisy_lena32} show the
original~256~shade images augmented with a low level of noise,
spanning $1/8$ of the total range~(32~shades of gray).  That is, the noise randomizes the 3 least significant bits (LSB)\index{least significant bit}\index{LSB} of the~8~bits in use.  If we now reduce to
to~8~shades (encodable by~3~bits) having added this noise, we get
\autoref{fig:random_8level_grayscale} and
\autoref{fig:random_8level_Lena}, which are a decided improvement on
\autoref{fig:FixedThreshold8} and \autoref{fig:FixedThresholdLena8},
which are what we get if we go from~256 shades to~8 shades without first
adding noise.

The process of adding noise to an image or a signal in order to preserve
information once the signal is subjected to
quantization\footnote{``Quantization'' here does not refer to the
physicist's process of finding a quantum\-/mechanical version of a
classical theory, but rather the process of discretizing the
properties of an image or
signal.}\index{quantization}~(digitization)\index{digitize}\index{digitization}\index{quantize}
is called \index{dithering}dithering.\footnote{Dithering also includes
related methods that use, not random noise, but some other signal which is
uncorrelated with the signal of interest. The ring laser gyroscope,
for example, uses periodic (sinusoidal) dither to prevent its
counter\-/rotating laser beams from locking under conditions of slow rotation.} In the example above, we took an
already discrete signal (each pixel having one of 256 shades of gray)
and made it more discrete, mapping the 256 shades into 2 shades (black
and white). However, the initial process of moving from an image with a
continuum of shades to one with 256 shades is also an example of
quantization. Were we dealing with digital audio, we would be working
with a one\-/dimensional stream of samples of the waveform, each of which
has a continuously valued amplitude~(the volume) that must be mapped
into a finite set of numbers for storage in a computer, say 24 bits
(16,777,216 possible values). This can then be further reduced to 16
bits (64,436 possible values) for CD encoding. Dither is routinely used
in this process.

Most digital representations involve more than one threshold. The
seminal work of \citet{Roberts62} considered the problem of transmitting
digital television. At the time, a 6-bit-per\-/pixel resolution was
considered adequate. \Citeauthor{Roberts62} proposed a scheme whereby a
3-bit-per\-/pixel signal could effectively encode the necessary detail if
pseudo\-/random noise were added to the 6-bit representation prior to
rounding to 3 bits.\footnote{The scheme of \protect\citet{Roberts62} was
an early example of what is called subtractive dither, where the noise
is subtracted from the image after transmission}. Shortly thereafter,
others realized that this technique was akin to a technique called
``dither'' that had been conceived several decades prior as a technique
to overcome the tendency of certain mechanical systems to stick for
various reasons, rendering them insensitive to small changes in
operational parameters~\citep{Schuchman64}. Engineers designed
electromotor circuits to apply dither in the form of small zero\-/mean
oscillations that would allow devices to respond more easily to small
steering signals from the operator~\citep{Farmer44, KK52}. In all of
these cases, proper application of dither tends to linearize a nonlinear
system; the dither blurs thresholds, eliminating some of the jaggedness
that goes along with systems that have one or more thresholds.

But the addition of noise does something else along the way. In a
physical system, adding noise adds energy to the system. If the system
has one or more thresholds, this has the effect of taking subthreshold
signals and boosting them, albeit stochastically. This is the phenomenon
known as \emph{stochastic resonance}.\index{stochastic resonance} Let's take a look at a simple
example, again using printed images, before moving on to the role
stochastic resonance can play in natural~-- including biological~--
phenomena.

\begin{figure}
    \subcaptionbox{\label{fig:PEpal}8 bit grayscale before thresholding (threshold marked)}
       {\includegraphics[width=\textwidth, frame]{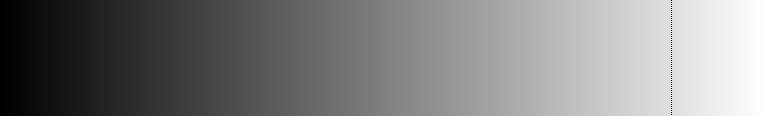}} 

    \subcaptionbox{\label{fig:PEpal_thresh}8 bit grayscale after thresholding}
      {\includegraphics[width=\textwidth, frame]{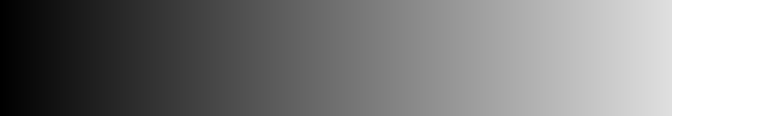}}

    \subcaptionbox{\label{fig:PE}8 bit signal before thresholding}
      {\includegraphics[width=\textwidth, frame]{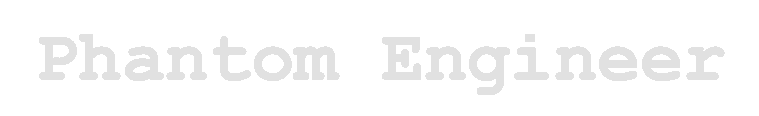}} 

     \subcaptionbox{\label{fig:PE_thresh}8 bit signal after thresholding}
        {\includegraphics[width=\textwidth, frame]{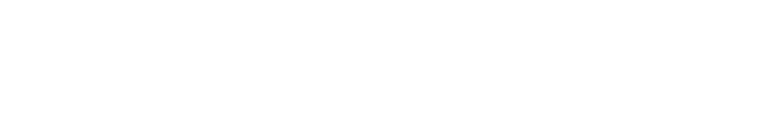}} 
    \caption{Rendering a signal imperceivable by thresholding. The signal is too light to survive the imposition of the threshold.}
    \label{fig:Phantom}
\end{figure}

We're accustomed to the fact that there are sounds we can't hear because
they're too soft, and sights we can't see because they're too faint.
These are thresholds of hearing and vision, respectively. By analogy,
consider an image we can't make out because it's too light: a very light
shade of gray indistinguishable from its white background.
\Autoref{fig:PEpal} shows the full grayscale palette with a line
separating the grays that are dark enough to distinguish from those
that, for some focal individual, are not. We can represent the indistinguishability of any grays below the threshold by rendering them as white, as in
\autoref{fig:PEpal_thresh}. Thus, when a faint image of the words `Phantom Engineer' (\autoref{fig:PE}) is
rendered in this very light gray, it will look like
\autoref{fig:PE_thresh} to someone for whom this threshold represents
the limits of their perceptual acuity. The image will be invisible.

\begin{figure}
     \subcaptionbox{\label{fig:NoisePE}Low-level (3 bit) noise before thresholding}
        {\includegraphics[width=\textwidth, frame]{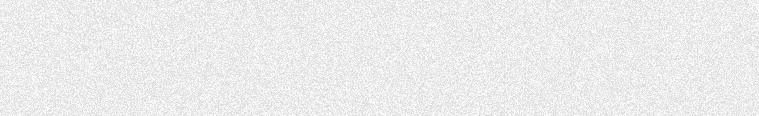} }

      \subcaptionbox{\label{fig:NoisePE_thresh}Low-level (3 bit) noise after thresholding}
        {\includegraphics[width=\textwidth, frame]{PE_white_noborder}} 

      \subcaptionbox{\label{fig:PE_noise}8 bit signal plus 3 bit noise before thresholding}
        {\includegraphics[width=\textwidth, frame]{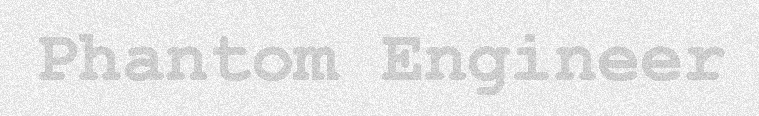}} 

      \subcaptionbox{\label{fig:PE_noise_thresh}8 bit signal plus 3 bit noise after thresholding}
        {\includegraphics[width=\textwidth, frame]{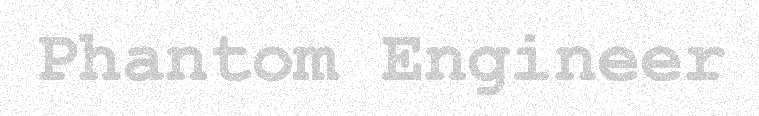}}
    \caption{Stochastic resonance: signal boosting with noise. Random dither noise spanning the 3 least significant bits (LSB) is added. With the addition of noise, the signal becomes dark enough to remain visible even after the imposition of the threshold.}
    \label{fig:NoisyPhantom}
\end{figure}

Now, suppose we add noise by randomly darkening each pixel,
including the background. We will use low\-/level,~3~bit noise, as shown to the right of the vertical bar in the grayscale palette depicted in
\autoref{fig:NoisePE}.  This corresponds to randomly selected shades of the very light grays
lying below the threshold of perceivability.\index{threshold of perceivability} Thus, like the image itself, the noise
will be invisible --~see \autoref{fig:NoisePE_thresh} --~to someone who
cannot make out very light shades of gray. Adding this noise to our
original image as in \autoref{fig:PE_noise} has the remarkable effect of
bringing a noisy but very legible version of the image above the
threshold of perceivability, as is evident in
\autoref{fig:PE_noise_thresh}, in which the lightest, sub-threshold grays are removed. The ability of noise to boost a signal above threshold is the essence
of stochastic resonance.\index{stochastic resonance}

One of the salient characteristics of stochastic resonance, indeed the
feature that makes it somewhat akin to a true resonance phenomenon, is
the dependence on the amplitude and specific properties of the noise.
This is true of dither noise in general. Too much noise (here, too large
a spectrum of grays) threatens to obliterate the signal, while too
little will fail to push the signal above threshold at all, and have no
effect. The relevance of this for understanding the role of stochastic
resonance in nature is significant, for there is noise of all kinds and
all amplitudes everywhere. Oftentimes it does what we think noise does:
it interferes with the signal, the image, or the operation of a
dynamical system. But when the noise level is proportional to the level
of one or more significant thresholds in a system of interest, we can
and should look for stochastic resonance, as it may be key to
understanding the function of the
system~\citep{gammaitoni1998stochastic}.

\section{Amplification by noise in natural dynamical systems}
\label{sec:stochres}

We will now take a look at how the stochastic resonance effect can be
used in modeling a dynamical system existing in nature. The example
we'll study is the one in which the term `stochastic resonance' was
originally introduced.  Though there is no resonance in the ordinary
physicists' sense~\citep[though see][]{gammaitoni1995stochastic}, we are
once again presented with a situation in which the addition of noise
permits the system of interest~-- in this case the climate~-- to
straddle a threshold.

\begin{figure}\centering
    \includegraphics[width=0.85\textwidth]{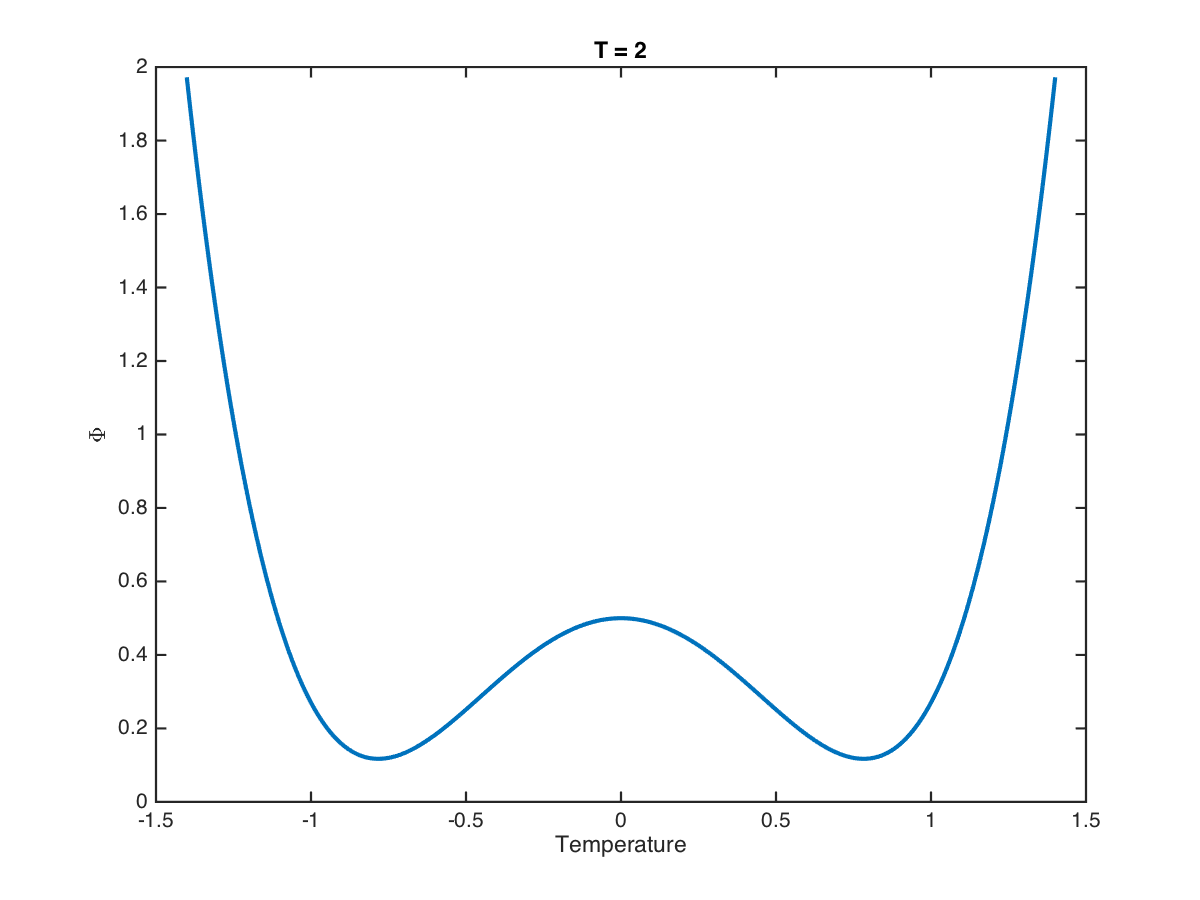}
    \caption{Double\-/well pseudo\-/potential representing the earth's
    climate as a dynamical system}
    \label{fig:Doublewell2}
\end{figure}

The earth has existed in \index{bistability}two relatively stable
climates around 10 degrees Kelvin apart for millions of years. Periods
in which the climate is cooler are called ice ages. In the late
1970s~\citep{BG78}, it was conjectured that the two stable climates
correspond to the two-minima of a double\-/well pseudo\-/potential like
the one shown in \autoref{fig:Doublewell2}.
The horizontal axis represents the earth's temperature, and the curve
acts like a potential energy term in ordinary mechanics, with a single
unstable equilibrium forming an energy barrier between
\index{bistability}two stable equilibria. In this model, the earth's
climate inevitably converges to one of the two stable equilibria. The
equilibrium on the left represents the ice age, and the one on the right
is a temperate period like the present.

Thus, we have a primitive model of a system with \index{bistability}two
stable states. But the stability of these states means that there is no way to transition between them, thus no way to explain how the climate shifts from one to the other. However, it was observed that the
eccentricity of the \index{earth's orbit}earth's orbit varies over a
period roughly the same as the time between ice ages~--~around~100,000
years~\citep{HIS76}. Because the eccentricity of the orbit is correlated
with small variations in the amount of solar
heating~(``\index{noise!solar heating|see{insolation}}%
\index{variation in solar heating|see{insolation}}%
\index{solar heating|see{insolation}}%
\index{noise!insolation|see{insolation}}%
\index{insolation}\emph{insolation}''),
it was conjectured that these small variations might be sufficient
external drivers of the earth's dynamical system to
cause the observed periodic \index{climate change}climate changes. This suggests that one augment the model by introducing a time\-/varying oscillation in the pseudo\-/potential in which the double\-/well shape is a transient
feature separating two epochs in which the potential morphs into a single well, a single quasi-stable
equilibrium.

\begin{figure}\centering
      \subcaptionbox{$t = 0$\label{dw1}}
        {\resizebox{!}{0.3\textheight}{\includegraphics[width=0.625\textwidth]{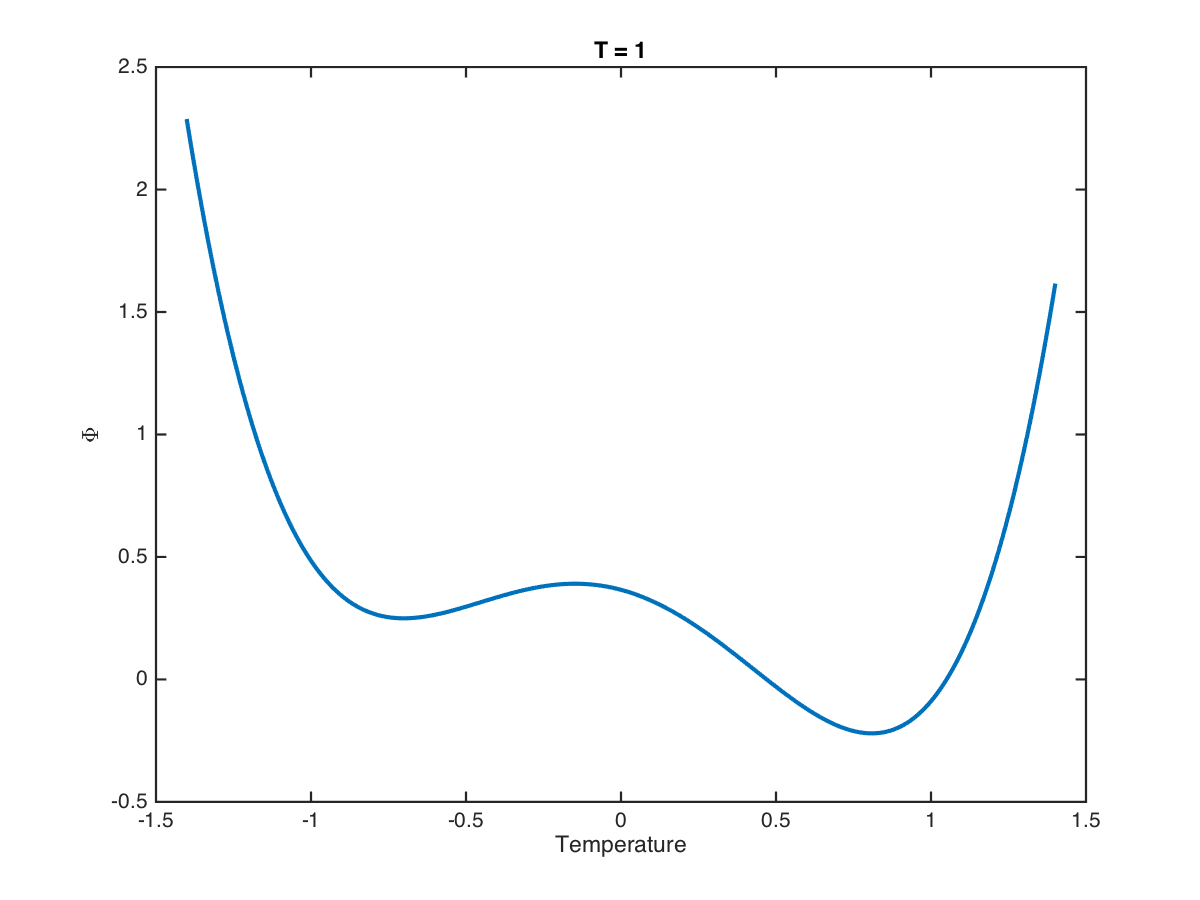}}}

      \subcaptionbox{$t = 25,000\,\text{years}$\label{dw2}}
        {\resizebox{!}{0.3\textheight}{\includegraphics[width=0.625\textwidth]{dw2}}}

      \subcaptionbox{$t = 50,000\,\text{years}$\label{dw3}}
        {\resizebox{!}{0.3\textheight}{\includegraphics[width=0.625\textwidth]{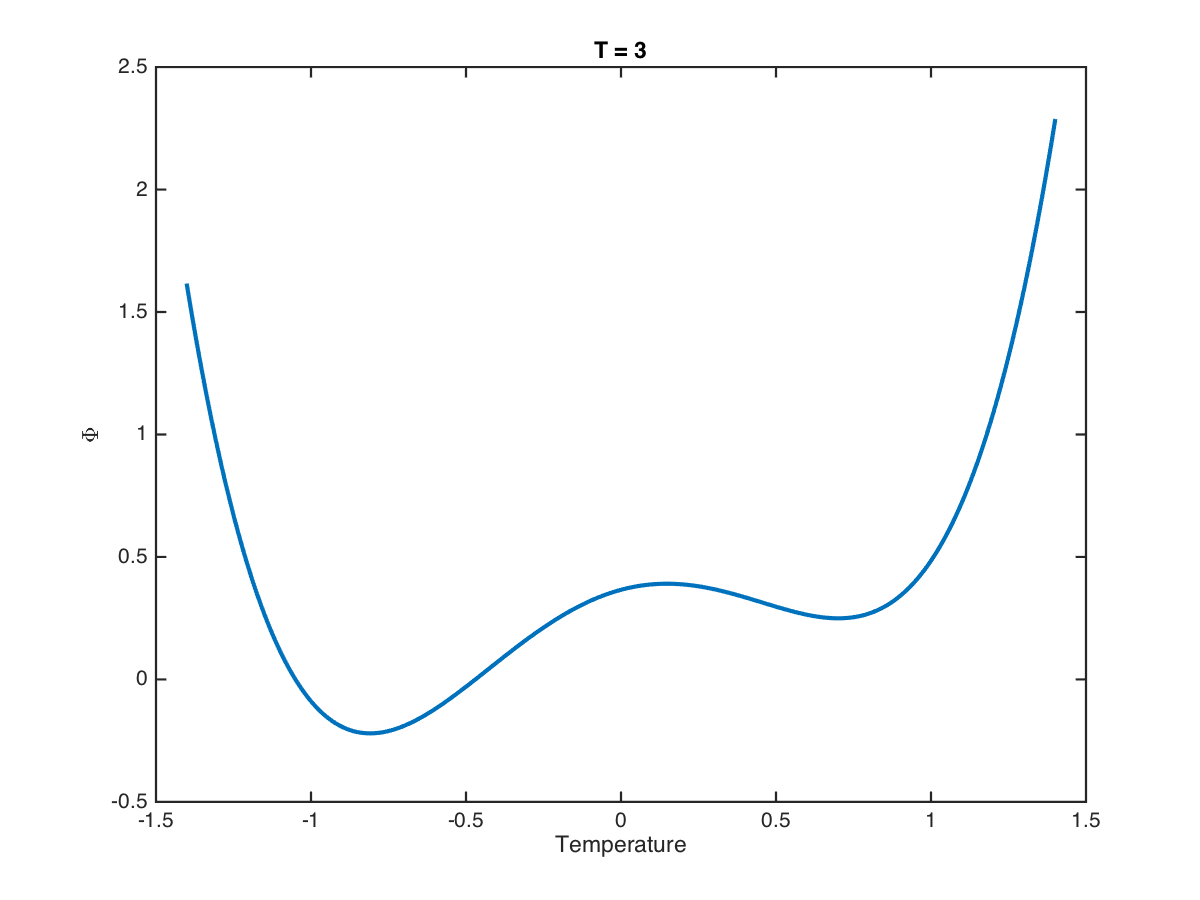}}}
      \caption{Time\-/varying double\-/well pseudo\-/potential with
      period of 100,000 years}\label{fig:dw}
\end{figure}

The problem with this idea was that the estimated insolation differences were too small to be
responsible for such a change. At best, the resulting time\-/varying
pseudopotential takes forms like those in \autoref{fig:dw}, where the
barrier between the \index{bistability}two stable equilibria is always
maintained, and where the change in temperature due to the displacement of each local minimum 
is of the order of only 1 degree Kelvin.
So a simple dynamical systems approach does not provide an explanation
for the congruence of the period of the insolation signal and the period
of earth's climate switching.

Independently, \citet{Nicolis82} and \citet{BPSV82} arrived at similar
explanations for how the climate might actually shift. \index{stochastic
resonance}They proposed that including the fine-scale, shorter-time
variations in heating and cooling due to various other factors might
result in the climate hopping from one well to the other.  In other
words, factoring in the existence of a certain level of noise in the
climate system might account for the ability, as it were, of the climate
to surmount the otherwise insurmountable threshold, the hump between the
two minima. After all, the geological record shows not only periodicity
in the earth's climate but also
significant small\-/scale variations which indicated that the earth's
dynamics must include some internal noise. So, following the approach of
\citeauthor{Nicolis82}, the deterministic double\-/well
pseudo\-/potential is augmented with a noise source, converting an
ordinary differential equation into a stochastic differential equation.
In other words, the climate is now modeled as a diffusion process~--~a
random walk that is pulled downhill but can, on occasion, take several
steps uphill. For such a diffusion process in a double\-/well, the mean time
to transition from one stable equilibrium to another is well
characterized by a formula parameterized by the height of the barrier
between the equilibria and the strength of the internal noise. What
\citeauthor{Nicolis82} realized was that the changes in the barrier
height due to noise could lead to
large changes in the mean residence time. If the noise is of the right
amplitude, then the climate is likely to hop from one not\-/quite\-/stable
minimum to the other when the barrier is low. Consequently,
\citet{BPSV82} named the phenomenon
\index{resonance!stochastic|see{stochastic resonance}}%
\index{stochastic resonance}%
\emph{stochastic resonance} based on its similarity to the
frequency\-/selective properties of conventional deterministic
resonance. Whereas ``resonance'' in the traditional sense is between the
frequency of an input and the characteristic response of a system, the
resonance here is between the frequency of the long-term oscillation in
insolation (the input) and the amplitude of the noise (a characteristic
feature of the system). If the noise is too small, nothing special will
happen; the system will never transit from one climate to the other. If
the noise is too great, the system will never settle in one climate or
another, as the noise will dominate the oscillation. If the noise is
within the correct range, however, the climate will oscillate at
approximately the 100,000 year period of the subthreshold background
oscillation in the ellipticity of the earth's orbit.

\section{Noise in living systems: decision making in ant colonies}
\label{sec:decisionmaking}

Over the past~20 years, an intriguing body of evidence has pointed to a
role for \index{stochastic resonance}stochastic resonance in a variety
of biological processes~\citep{MA09}. Extensive work has been done
demonstrating the role of noise in general and stochastic resonance in
particular in the neural systems that carry out sensory information
processing~\citep{moss2004stochastic}, but it is important at the
macroscopic level as well. We will conclude our discussion of
stochastic resonance by discussing its role in the social dynamics of
group decision making in certain species of ants.

There are a wide variety of \index{mass\-/recruiting ants|see{ants,
mass\-/recruiting}}
\emph{mass\-/recruiting ants} that form charismatic foraging trails that
concentrate all foraging effort onto a single food source for a short
period of time~\citep{HW90}. As many species of mass\-/recruiting ant
have a heterogeneous foraging force, it is thought that it may be
beneficial to concentrate the foraging force all in one area in order to
guarantee there is an adequate representation of each worker type. So it
is expected that these ants must make use of some decentralized
mechanism that can drive its foraging force to a quick consensus on the
best of several available foraging options.
\index{trail\-/laying ants|see{ants, mass\-/recruiting}}%
\index{ants!mass\-/recruiting}A typical feature of
mass\-/recruiting ants is the use of \index{pheromone trails|see{ants,
pheromone trails}}%
\index{ants!pheromone trails}pheromone trails~\citep{HW90}. Although
details vary across different mass\-/recruiting ant taxa, the observed
pattern is usually a variation of what follows. A focal ant leaves her
nest and searches for food. When she finds food, she can choose to
return to her nest and deposit some quantity of \index{ants!pheromone
trails}pheromone along
her path back to the nest. The amount of \index{ants!pheromone trails}%
pheromone she deposits is related to the quality of the discovered food,
with higher\-/quality foods leading to more deposited
\index{ants!pheromone trails}pheromone. Although that deposited
\index{ants!pheromone trails}pheromone will eventually evaporate, for
a short time after deposition, the \index{ants!pheromone trails}pheromone
near the nest will attract the attention of other foragers that would
otherwise search randomly for food. They will then have an increased
likelihood of finding the same food source as the focal ant and then
also lay a \index{ants!pheromone trails}pheromone trail on their return
visit. So initially, a set of food items will be discovered randomly.
Due to the positive feedback inherent in the recruitment system, the
highest quality of those food sources will eventually attract all of the
foragers.

\begin{figure}\centering
    \includegraphics[width=\columnwidth]{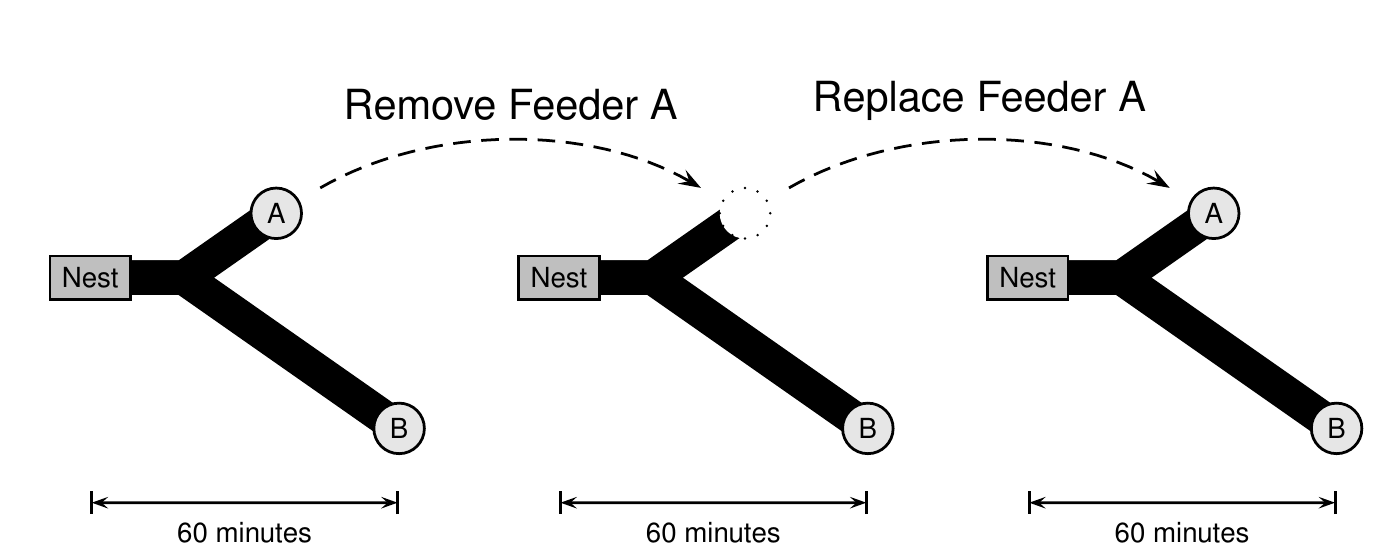}
    \caption
        [Dynamic foraging experiment with \emph{Pheidole megacephala}
         ant colonies]
        {Graphical summary of dynamic foraging experiment used by
         \protect\citet{DBNM09} to study flexibility of trail\-/laying
         in \emph{Pheidole megacephala} big\-/headed ant colonies. In
         the first 60 minutes of the experiment, colonies are given a
         choice between two feeders, A and B, that only differ in
         distance to the nest. During the second 60 minutes of the
         experiment, the nearest feeder~(A) is removed. Finally, during
         the final 60 minutes of the experiment, the nearest feeder~(A)
         is replaced.}
    \label{fig:DBNM09}
\end{figure}

Until recently, it has been believed that such
\index{ants!mass\-/recruiting}trail\-/laying mass\-/recruitment
mechanisms had a flaw similar to the one described in
\autoref{sec:stochres} for the early deterministic models of
\index{climate change}climate change~--~the ants were thought to be
rigidly \index{bistability}bistable and unable to cope with changes in
food availability after a critical point in the recruitment process. In
other words, the positive feedback in the recruitment would eventually
become so strong that the system would become entirely insensitive to
changes in relative food-source quality, just as early mathematical
models of \index{climate change}climate change were not properly
sensitive to the variations in \index{insolation}insolation. This
intuition was verified in early experiments with
\index{Lasius@\emph{Lasius}}\emph{Lasius niger}~\citep{BDGP90}.
Moreover, early dynamical mathematical models of trail\-/laying were
also shown to be insensitive to changes in relative food
quality~\citep{ND99, CDFSTB01}. However, several recent experiments show
that many other \index{ants!mass\-/recruiting}trail\-/laying ants are
able to dynamically re\-/allocate their foraging forces to track changes
in the environment~\citep{DBNM09, RSB11, LB13}. For example,
\citet{DBNM09} presented colonies of
\index{Pheidole@\emph{Pheidole}}\emph{Pheidole megacephala} with a
laboratory dynamic environment summarized in \autoref{fig:DBNM09}.
During the 180\-/minute experiment, colonies were placed at
the mouth of a Y-bridge with two legs of different lengths, and the
experiment proceeded in three 60\-/minute phases:
\begin{itemize}

    \item During the first~60 minutes, equal\-/quality feeders were
        placed at the ends of both legs. Because one leg was shorter, it
        eventually dominated the collective attention of the colony and
        a single trail was formed to the feeder on that leg.

    \item During the second~60 minutes of the experiment, the feeder on
        the short leg was removed. With the disappearance of food, the
        pheromone trail was not reinforced, and the colony was
        eventually able to return to random search and subsequently
        converge on the feeder at the end of the long leg.

    \item During the final~60 minutes of the experiment, a feeder was
        returned to the short leg of the Y-maze. The traditional model
        of trail\-/laying recruitment would predict that the new feeder
        would be ignored because all foragers would be latched into
        following the existing pheromone trail. However, contrary to
        those predictions, the short leg was re\-/discovered and ants
        returned to exploiting the closer feeder.

\end{itemize}

To explain the results of the experiment, \citeauthor{DBNM09} propose a slight extension
to the traditional mathematical model of trail\-/laying recruitment inspired by \index{stochastic resonance}stochastic resonance. They
observed that ants in their experiments would often make ``errors'' in
their \index{ants!mass\-/recruiting}trail\-/following behavior that
would lead a minority of the ants down the opposite leg of the Y-maze.
In an attempt to capture this phenomenon, they augmented the traditional
mathematical model of trail\-/laying behavior with an ``error'' level
that would cause an individual to rarely, but measurably often, choose
the leg of the Y-maze with the smaller quantity of pheromone deposited
on it. At small error levels, the theoretical system had nearly
identical decision\-/making latency and accuracy characteristics to the
deterministic system when presented with a static choice set. However,
at specific non\-/zero error levels, the system could produce switching
dynamics that matched those of experimental data from ants like
\emph{P.~megacephala} that have the ability to follow changes in
relative feeder quality. Furthermore, the mathematical model predicted
that different error levels would correspond to different random
natural switching times between alternatives, and amplification of
variations in food quality would be possible if the periodicity of those
variations matched the natural error\-/driven switching time.

The time\-/scale matching argument for the switching behavior observed
in some trail\-/laying, mass\-/recruiting ants is identical to the one
used in the early models of stochastic resonance in climate systems
that are matched to the periodicity of solar insolation. However, in the
case of the ants, differences in error level across different ant
taxa could be explained by natural selection. In particular, the
individual error level could be tuned by natural selection so that the
stochastic switching time of the colony would match the natural
periodicity of changes in food quality in the natural environment.
Colonies with individuals that make errors at the appropriate rate would
have an advantage over colonies that make more or fewer errors. Making
too many individual\-/level errors would mean switching too frequently
from good choices to bad choices, and making too few individual\-/level
errors would mean focusing for too long on one choice even though
a better choice was now available. Consequently, the ``errors'' at
the individual level would be better described as random
variability~(noise) that was itself a trait under selection, and the
prediction would be that ants evolved in more ephemeral environments
would also have larger amounts of noise in individual\-/level response
to pheromone trails. In fact, as \citet{DBNM09} discuss, the flexible
\emph{P.~megacephala} ants in their study that are well modeled by
non\-/zero noise do come from an environment where food quality changes
more frequently than the \emph{L.~niger} ants that had previously been
used to support the deterministic modeling of trail\-/following behavior
with no noise. If location of the best\-/quality food source is viewed
as a signal that tends to change over some characteristic time scale, then the current location of the
main foraging trail can be viewed as a version of that signal amplified
using \index{stochastic resonance}stochastic resonance. This argument is
identical to those made in the \index{stochastic
resonance}stochastic\-/resonance literature where some
input\-/to\-/output measure, such as \index{information theory}%
\index{mutual information}mutual information, is maximized by varying
the amount of noise added to the input signal~\citep{NSAESF96}.
Similarly, in the earlier image processing example, a certain amount of
noise is sufficient to push the text `Phantom Engineer' above the
threshold of visibility~(\autoref{fig:NoisyPhantom}). Too little noise
will not do the trick. Adding noise will make the text more visible up to
a point, after which the readability goes down as the entire image
becomes dominated by noise. In the case of the ants, the signal being
amplified is the relative quality of the feeders, and the output is the
selection of a path by the colony.

The idea that apparent ``errors'' could actually be an adaptive
phenotypic trait under selection is not unprecedented and goes beyond the
examples of possible stochastic resonance in natural phenomena. For
example, the idea that noise can be of benefit in
decision making is relatively old. For example,in what they called
\index{cybernetics}\index{ethological cybernetics}``ethological
cybernetics,'' \citet{HS54} performed an \index{information theory}%
information\-/theoretic analysis of the statistical distributions
of \index{honeybees}\index{Apis@\emph{Apis}|see{honeybees}}honeybees
responding to communicated information from so\-/called ``waggle
dances.'' Honeybees have the ability to communicate information from one
forager to another through a dance language that communicates the
relative polar coordinates (i.e., distance and direction) of a
discovered food source. However, after a bee communicates these
coordinates to another, the bee receiving the information will often
make ``mistakes'' and explore a location slightly different from the one
discovered by the original bee. The \emph{average} location explored by an
ensemble of receiver bees will closely match the originally discovered
food source, and so these variations are viewed as ``noise'' due to
imperfect communication of the coordinates in the
dance\-/communication channel. \Citeauthor{HS54}
determine that the bee-to-bee channel communicates roughly 4 bits of
information about the direction of the target. That is, a bee can only
communicate 1 of 16 different cardinal directions; any finer resolution
appears to be impossible. This may seem to reflect a fundamental
limitation, such as a physiological or neurological constraint, but it
could also be an adaptive response to dispersed food sources in an
environment. If a \index{honeybees}honeybee is dancing to communicate
the location of a nectar source, such as a flower, the resulting noisy
scatter of her colony mates will likely find other flowers in a similar
location. Thus, the amount of error in the communication may be tied to
some ecological measure of forage patchiness, and honeybees selected for
environments with a different patchiness may communicate with different
levels of error.

As the honeybee example does not involve dynamically changing signals in
the environment, it is not an example of stochastic resonance in the strict sense, but it does reflect how the amount of noise expressed in a behavior is itself a phenotype that nature can adapt to match natural variation. However, a very similar information\-/theoretic analysis of fire ants does suggest additional
ties to stochastic resonance and behaviors shaped by nature. In
particular, \citet{Wilson62b} described a consistent error distribution
in the distance and direction information communicated by
\index{ants!fire|see{fire ants}}\index{fire ants}%
\index{Solenopsis@\emph{Solenopsis}|see{fire ants}}fire\-/ant
\index{ants!pheromone trails}pheromone trails leading to prey. In
the trail\-/laying examples above, the actual food sources were static,
and the experimenters could change the location of different sources at
discrete instants of time. This was appropriate for the particular ant species under study. 
However, fire ants are a natural example of a species adapted to continuously varying food quality and location. These ants track moving food sources: living prey items that have the ability to flee. They must have the ability to dynamically adapt and follow fleeing prey until the prey is sufficiently subdued.
\Citeauthor{Wilson62b} suggests that the relatively poor ability of
individual fire ants to follow trails is actually an adaptation. The result of the ensemble of error\-/prone trail
followers is a cloud of ants in the general vicinity of the original
location of the discovered prey. If this cloud is large enough, it can
track the motion of the escaping prey. Too much noise will cast too
large of a net and lead to too thin coverage over a prey item, and too
little noise will not disperse the ants far enough to catch the escaping
prey. So the dispersal of the trail followers could be matched to the
escape dynamics of the typical kinds of prey. Just as in traditional
stochastic resonance examples, a certain critical amount of noise helps a dynamic
output~(the ultimate location of the end of a fire\-/ant foraging
trail) follow a dynamic input~(the trajectory of an escaping prey
item).

\section{Conclusion}
\label{sec:conclusion}

An ant colony's ability to react to changes in the food supply and the
climate's ability to react to small changes in insolation are examples
of the power of noise to qualitatively change the way a system responds
to its environment. If we disregard the noise~-- disregard the small,
random variations in the properties of the system~-- we find the system
converges to a fixed point of its dynamical equations and stays there
indefinitely. The ant colony remains fixated on a single food source,
insensitive to changes in food supply; the climate remains where it is,
never shifting. But when the model of the system is modified to
include noise, the system is able to surmount a barrier and transition
to a qualitatively distinct state. The colony is able to discover and
consolidate a new path. The climate reacts to the slight change in
insolation over the course of millenia. Noise makes these systems more
sensitive to their environment.

A promising area to look for further noise effects in biology lies at
smaller scales. The fundamental process at the foundation of all life is
the expression of genes as proteins. The chemical reactions involved are
constrained by both the availability of the reactants and their
proximity, making the process as a whole subject to fluctuations which
have come to be called \emph{gene expression
noise}~\citep{KEBC05}\index{gene~expression~noise}. The sources of the
noise and the role it plays in the process of development and
reproduction are matters of intense contemporary investigation
\citep{sanchez2013regulation, Viney20131104}.\index{phenotypic~noise}
For example, \citet{fernando2009molecular} proposed the existence of an
intracellular genetic perceptron, a single cell gene network capable of
associative learning.\index{genetic perceptron}\index{associative
learning} Remarkably,\Citet{bates2014stochastic} show that the performance of the
perceptron is actually enhanced by gene expression noise at a specific level.

One of the factors impacting gene expression in bacteria is the
intercellular communication mechanism broadly known as \emph{quorum
sensing}\index{quorum sensing}, in which bacteria both emit and detect
signaling molecules, allowing them to infer the concentration of other
bacteria and act accordingly~\citep{waters2005quorum,
popat2015collective}. This, too, is a noisy process, subject to the
whims of diffusion in the intercellular environment. Like the
examples of stochastic resonance we have considered, it is a
threshold\-/oriented system, whereby genes are switched on or
off depending on whether a critical density of other bacteria are sensed
in the neighborhood. \Citet{karig2011model} have applied stochastic
resonance to the development of a synthetic biological system which
utilizes gene expression noise to boost a time\-/varying molecular signal
consisting of varying concentrations of the molecule used by
Gram\-/negative bacteria in quorum sensing.\index{synthetic biology}

The recent, fascinating work on the role in biological systems of
stochastic resonance in particular, and noise in general, is surely the
tip of an iceberg. \Citet{hoffmann2012life} advances the idea that the
molecular machines like kinesin that do physical work within the cell make use of
the random noise that is the thermal motion of the water molecules in
the cytoplasm.\index{molecular ratchet} \Citet{rolls2010noisy} provide
an extended look at the role of noise in brain function\index{brain}.
The idea that noise can and does do work, enhancing the information
processing that is essential to life, is an idea whose time has come.


\bibliography{Noise3_arXiv}\label{refs}

\begin{thebibliography}{38}
\expandafter\ifx\csname natexlab\endcsname\relax\def\natexlab#1{#1}\fi
\expandafter\ifx\csname selectlanguage\endcsname\relax
  \def\selectlanguage#1{\relax}\fi

\bibitem[\protect\citename{Bates {et~al.}, }2014]{bates2014stochastic}
Bates, Russell, Blyuss, Oleg, and Zaikin, Alexey. 2014.
\newblock Stochastic resonance in an intracellular genetic perceptron.
\newblock {\em Physical Review E}, {\bf 89}(3), 032716.

\bibitem[\protect\citename{Beckers {et~al.}, }1990]{BDGP90}
Beckers, R., Deneubourg, Jean-Louis, Goss, Simon, and Pasteels, Jacques~M.
  1990.
\newblock Collective decision making through food recruitment.
\newblock {\em Insectes Sociaux}, {\bf 37}(3), 258--267.

\bibitem[\protect\citename{Benzi {et~al.}, }1982]{BPSV82}
Benzi, Roberto, Parisi, G., Sutera, Alfonso, and Vulpiani, Angelo. 1982.
\newblock Stochastic resonance in climate change.
\newblock {\em Tellus}, {\bf 34}, 10--16.

\bibitem[\protect\citename{Bhattacharya and Ghil, }1978]{BG78}
Bhattacharya, K., and Ghil, M. 1978.
\newblock An energy-balance model with multiply-periodic and quasi-chaotic free
  oscillations.
\newblock {Pages  299--310 of:} {\em Evolution of planetary atmospheres and
  climatology of the earth}.
\newblock Toulouse, France: Centre National dÕEtudes Spatiales.

\bibitem[\protect\citename{Camazine {et~al.}, }2001]{CDFSTB01}
Camazine, Scott, Deneubourg, Jean-Louis, Franks, Nigel~R., Sneyd, James,
  Theraulaz, Guy, and Bonabeau, Eric. 2001.
\newblock {\em Self-Organization in Biological Systems}.
\newblock Princeton, NJ, USA: Princeton University Press.

\bibitem[\protect\citename{Dobrindt and Hacker, }2001]{dobrindt2001whole}
Dobrindt, Ulrich, and Hacker, J{\"o}rg. 2001.
\newblock Whole genome plasticity in pathogenic bacteria.
\newblock {\em Current opinion in microbiology}, {\bf 4}(5), 550--557.

\bibitem[\protect\citename{Dussutour {et~al.}, }2009]{DBNM09}
Dussutour, Audrey, Beekman, Madeleine, Nicolis, Stamatios~C., and Meyer, Bernd.
  2009.
\newblock Noise improves collective decision-making by ants in dynamic
  environments.
\newblock {\em Proc.\ R.\ Soc.\ B}, {\bf 276}(1677), 4353--4361.

\bibitem[\protect\citename{Farmer, }1944]{Farmer44}
Farmer, William~C. (ed). 1944.
\newblock {\em Ordnance Field Guide: Restricted}.
\newblock Military Service Publishing Company.

\bibitem[\protect\citename{Fernando {et~al.}, }2009]{fernando2009molecular}
Fernando, Chrisantha~T, Liekens, Anthony~ML, Bingle, Lewis~EH, Beck, Christian,
  Lenser, Thorsten, Stekel, Dov~J, and Rowe, Jonathan~E. 2009.
\newblock Molecular circuits for associative learning in single-celled
  organisms.
\newblock {\em Journal of the Royal Society Interface}, {\bf 6}(34), 463--469.

\bibitem[\protect\citename{Floyd and Steinberg, }1976]{FS76}
Floyd, Robert~W., and Steinberg, Louis. 1976.
\newblock An adaptive algorithm for spatial grey scale.
\newblock {\em Proc.\ Soc.\ Inf.\ Disp.}, {\bf 17}, 75--77.

\bibitem[\protect\citename{Fraser and Kaern, }2009]{fraser2009chance}
Fraser, Dawn, and Kaern, Mads. 2009.
\newblock A chance at survival: gene expression noise and phenotypic
  diversification strategies.
\newblock {\em Molecular microbiology}, {\bf 71}(6), 1333--1340.

\bibitem[\protect\citename{Gammaitoni {et~al.},
  }1995]{gammaitoni1995stochastic}
Gammaitoni, L, Marchesoni, F, and Santucci, S. 1995.
\newblock Stochastic resonance as a bona fide resonance.
\newblock {\em Physical review letters}, {\bf 74}(7), 1052.

\bibitem[\protect\citename{Gammaitoni {et~al.},
  }1998]{gammaitoni1998stochastic}
Gammaitoni, Luca, H{\"a}nggi, Peter, Jung, Peter, and Marchesoni, Fabio. 1998.
\newblock Stochastic resonance.
\newblock {\em Reviews of modern physics}, {\bf 70}(1), 223.

\bibitem[\protect\citename{Haldane and Spurway, }1954]{HS54}
Haldane, John B.~S., and Spurway, H. 1954.
\newblock A statistical analysis of communication in
  ``{\emph{\MakeUppercase{a}pis mellifera}}'' and a comparison with
  communication in other animals.
\newblock {\em Insectes Sociaux}, {\bf 1}(3), 247--283.

\bibitem[\protect\citename{Hays {et~al.}, }1976]{HIS76}
Hays, J.~D., Imbrie, John, and Shackleton, N.~J. 1976.
\newblock Variation in the earth's orbit: pacemaker of the ages.
\newblock {\em Science}, {\bf 194}(4270), 1121--1132.

\bibitem[\protect\citename{Hoffmann, }2012]{hoffmann2012life}
Hoffmann, Peter~M. 2012.
\newblock {\em Life's ratchet: how molecular machines extract order from
  chaos}.
\newblock Basic Books.

\bibitem[\protect\citename{H{\"{o}}lldobler and Wilson, }1990]{HW90}
H{\"{o}}lldobler, Bert, and Wilson, Edward~O. 1990.
\newblock {\em The Ants}.
\newblock Harvard University Press.

\bibitem[\protect\citename{Hutchison, }2001]{Hutchison01}
Hutchison, Jamie. 2001.
\newblock Culture, communication, and an information age
  {\MakeUppercase{m}}adonna.
\newblock {\em {IEEE} Prof.\ Commun.\ Soc.\ Newsl.}, {\bf 45}(3), 1, 5--7.

\bibitem[\protect\citename{K{\ae}rn {et~al.}, }2005]{KEBC05}
K{\ae}rn, Mads, Elston, Timothy~C., Blake, William~J., and Collins, James~J.
  2005.
\newblock Stochasticity in gene expression: from genotypes to phenotypes.
\newblock {\em Nat.\ Rev.\ Genet.}, {\bf 6}(June), 451--464.

\bibitem[\protect\citename{Karig {et~al.}, }2011]{karig2011model}
Karig, David~K, Siuti, Piro, Dar, Roy~D, Retterer, Scott~T, Doktycz, Mitchel~J,
  and Simpson, Michael~L. 2011.
\newblock Model for biological communication in a nanofabricated cell-mimic
  driven by stochastic resonance.
\newblock {\em Nano communication networks}, {\bf 2}(1), 39--49.

\bibitem[\protect\citename{Korn and Korn, }1952]{KK52}
Korn, Granino~Arthur, and Korn, Theresa~M. (eds). 1952.
\newblock {\em Electronic Analog Computers: {\MakeUppercase{D}}-c Analog
  Computers}.
\newblock McGraw-Hill.

\bibitem[\protect\citename{Latty and Beekman, }2013]{LB13}
Latty, Tanya, and Beekman, Madeleine. 2013.
\newblock Keeping track of changes: the performance ofa nt colonies in dynamic
  environments.
\newblock {\em Anim.\ Behav.}, {\bf 85}(3), 637--643.

\bibitem[\protect\citename{McDonnell and Abbott, }2009]{MA09}
McDonnell, Mark~D., and Abbott, Derek. 2009.
\newblock What is stochastic resonance? Definitions, misconceptions, debates,
  and its relevance to biology.
\newblock {\em PLoS Comput.\ Biol.}, {\bf 5}(5), e1000348.

\bibitem[\protect\citename{Moss {et~al.}, }2004]{moss2004stochastic}
Moss, Frank, Ward, Lawrence~M, and Sannita, Walter~G. 2004.
\newblock Stochastic resonance and sensory information processing: a tutorial
  and review of application.
\newblock {\em Clinical neurophysiology}, {\bf 115}(2), 267--281.

\bibitem[\protect\citename{Neiman {et~al.}, }1996]{NSAESF96}
Neiman, Alexander, Shulgin, Boris, Anishchenko, Vadim, Ebeling, Werner,
  Schimansky-Geier, Lutz, , and Freund, Jan. 1996.
\newblock Dynamical entropies applied to stochastic resonance.
\newblock {\em Phys.\ Rev.\ Lett.}, {\bf 76}(23), 4299--4302.

\bibitem[\protect\citename{Nicolis, }1982]{Nicolis82}
Nicolis, C. 1982.
\newblock Stochastic aspects of climatic transitions---response to a periodic
  forcing.
\newblock {\em Tellus}, {\bf 14}, 1--9.

\bibitem[\protect\citename{Nicolis and Deneubourg, }1999]{ND99}
Nicolis, Stamatios~C., and Deneubourg, Jean-Louis. 1999.
\newblock Emerging patterns and food recruitment in ants: an analytical study.
\newblock {\em J.\ Theor.\ Biol.}, {\bf 198}, 575--592.

\bibitem[\protect\citename{Penzias and Wilson, }1965]{PW65}
Penzias, A.~A., and Wilson, R.~W. 1965.
\newblock A measurement of excess antenna temperature at 4080
  {\MakeUppercase{m}}c/s.
\newblock {\em Astrophys.\ J.}, {\bf 142}, 419--421.

\bibitem[\protect\citename{Popat {et~al.}, }2015]{popat2015collective}
Popat, R, Cornforth, DM, McNally, L, and Brown, SP. 2015.
\newblock Collective sensing and collective responses in quorum-sensing
  bacteria.
\newblock {\em Journal of The Royal Society Interface}, {\bf 12}(103),
  20140882.

\bibitem[\protect\citename{Reid {et~al.}, }2011]{RSB11}
Reid, Chris~R., Sumpter, David J.~T., and Beekman, Madeleine. 2011.
\newblock Optimisation in a natural system: {\MakeUppercase{a}}Rgentine ants
  solve the {\MakeUppercase{t}}owers of {\MakeUppercase{h}}anoi.
\newblock {\em J.\ Exp.\ Biol.}, {\bf 214}(January~1,), 50--58.

\bibitem[\protect\citename{Roberts, }1962]{Roberts62}
Roberts, Lawrence~Gilman. 1962.
\newblock Picture coding using pseudo-random noise.
\newblock {\em {IRE} Trans.\ Inf.\ Theory}, {\bf 8}(2), 145--154.

\bibitem[\protect\citename{Rolls and Deco, }2010]{rolls2010noisy}
Rolls, ET, and Deco, G. 2010.
\newblock The noisy brain.
\newblock {\em Stochastic dynamics as a principle of brain function.(Oxford
  Univ. Press, UK, 2010)}.

\bibitem[\protect\citename{Sanchez {et~al.}, }2013]{sanchez2013regulation}
Sanchez, Alvaro, Choubey, Sandeep, and Kondev, Jane. 2013.
\newblock Regulation of noise in gene expression.
\newblock {\em Annual review of biophysics}, {\bf 42}, 469--491.

\bibitem[\protect\citename{Schuchman, }1964]{Schuchman64}
Schuchman, Leonard. 1964.
\newblock Dither signals and their effect on quantization noise.
\newblock {\em {IEEE} Trans.\ Commun.\ Technol.}, {\bf 12}(4), 162--165.

\bibitem[\protect\citename{Viney and Reece, }2013]{Viney20131104}
Viney, Mark, and Reece, Sarah~E. 2013.
\newblock Adaptive noise.
\newblock {\em Proceedings of the Royal Society of London B: Biological
  Sciences}, {\bf 280}(1767).

\bibitem[\protect\citename{Wagner, }2014]{wagner2014arrival}
Wagner, Andreas. 2014.
\newblock {\em Arrival of the Fittest: Solving Evolution's Greatest Puzzle}.
\newblock Penguin.

\bibitem[\protect\citename{Waters and Bassler, }2005]{waters2005quorum}
Waters, Christopher~M, and Bassler, Bonnie~L. 2005.
\newblock Quorum sensing: cell-to-cell communication in bacteria.
\newblock {\em Annu. Rev. Cell Dev. Biol.}, {\bf 21}, 319--346.

\bibitem[\protect\citename{Wilson, }1962]{Wilson62b}
Wilson, Edward~O. 1962.
\newblock Chemical communication among workers of the fire ant
  {\textit{\MakeUppercase{s}olenopsis saevissima}}
  ({\MakeUppercase{f}r.~\MakeUppercase{s}mith}): 2.~{\MakeUppercase{a}}n
  information analysis of the odour trail.
\newblock {\em Anim.\ Behav.}, {\bf 10}(1--2), 148--158.

\end{thebibliography}
\bibliographystyle{cambridgeauthordate}

\cleardoublepage

\printindex

\end{document}